\documentclass[a4paper]{article}
\usepackage{epsfig}

\usepackage{amsfonts}
\usepackage{bbm}

\begin{document}


\title{Into the Fold:\\ Searching for a Theory of Natural Inference}\author{Thomas Marlow \\ \emph{School of Mathematical Sciences, University of Nottingham,}\\
\emph{UK, NG7 2RD}}

\maketitle

\begin{abstract}
We introduce relationalism and discuss how it is useful for interpreting probability theory and quantum mechanics. This paper is written in relatively lay terms and presumes no prior knowledge of quantum theory.
\end{abstract}

\textbf{Keywords}: Relationalism; Quantum Theory; Probability Theory.\\

In this short note we will try and learn a few things from the major modern physical theories of our time and then we shall try to apply these lessons to quantum theory.  First we will introduce a broad range of theories that we call `relational' theories, noting that quantum theory is not one of them.  Then we will discuss a possible way of making quantum theory relational as well.

Leibniz, a contemporary of Newton, introduced two principles which together form an approach to physics which is called `relationalism':

\begin{enumerate}

\item \emph{The Principle of Sufficient Reason (PSR)}:  Every single element of a physical theory must be rationally justified.

\item \emph{The Principle of Identifying the Indiscernible (PII)}:  If we have no rational reason or ability to distinguish features of a theory then we should identify them, else the PSR is not satisfied.

\end{enumerate}

\noindent One can consider these two principles criteria which ensure `rationality'---or, in the least, criteria which ensure a pragmatic approach \cite{Smolin05}.  They are basically principles that ensure that we adopt what might be called `parsimony'---they ensure that we only use facts or distinctions that we are rationally compelled to use.  It is pragmatic to adopt these two principles simply because there are clearly many more things that we cannot rationally justify in comparison to those things that we can, and rather than adopt a `trial-by-error' approach it seems more parsimonious to only use axioms or facts we can all accept by using our reasoning skills.  The PSR basically says that if we all sit back in our armchairs, mull it over for a while, and still cannot all agree that some fact about the world must be the case (without resorting to mere proclamations of faith), then we must reject it from our physical theory.   (Of course, it might be that what we think as reasonable and rational now might turn out to be irrational in the future, but this should not stop us trying our best.)

There is only one argument we know of against relationalism when framed in such pragmatic terms; namely that it is sometimes \emph{even more pragmatic} to use arbitrary conventions which we cannot justify rationally, simply to make the maths easier.  However, relationalism, in some form or another, helped people to design all the modern physical theories we investigate today (with the exception of quantum theory).  For example, relationalism helped Einstein to design General Relativity (GR).  Einstein ensured that he only ever discussed the relationships between events---`spacetime coincidences'---and he tried not to add anything spurious to the catalogue of spacetime coincidences because these were the only elements in the theory he could rationally justify (by the PSR he did not want to use anything \emph{other} than the catalogue of spacetime coincidences).  Clearly we don't want to, by convention alone, assign numbers arbitrarily to all objects in the universe and then call these numbers `positions'.  This is exactly what we \emph{would} be doing if we were to try and define positions absolutely.  We simply don't have a device---one might call it a `positometer'---which, when we point it at an object (and that object alone), reads off the object's `position' \cite{unknown}.  Similarly, if we look down our best microscopes we do not see magical glowing numbers that we might call positions.  Clearly we do have devices that can read off positions \emph{if we make arbitrary conventions} such as ``distances and angles are measured relative to the Eiffel tower''---this is not what we mean by a positometer.  Anyone can make arbitrary conventions but we want to find physical theories that aliens could also discover.  If we wish to enter communion with such aliens---or, to put it differently, with the natural world itself---we must not use such arbitrary human conventions.  It turns out that such an approach actually \emph{helps} us find physical theories.  This is borne out by the fact that the essential features of all modern physical theories, except one, are relational in this `parsimonious' sense.  So we must reject the idea of a positometer by the PSR.  This rejection, via Einstein and a lot of hard work, gave us GR, a relational approach to spacetime.

As Smolin has emphasised \cite{SmolinBOOK2}, it turns out that all modern gauge theories (which includes the standard model and string theory) also have relational aspects.  Gauge theories, excluding GR, do not treat spacetime relationally (although of course they might in the future) but they do treat charge conventions relationally.  Just as we don't have a positometer, we don't have devices which measure what absolute sign a charge, like an electron, has.  Clearly we do have devices which can measure the signs of charges if we make arbitrary conventions such as ``the electrons contained within Tony Blair are all negative'' but there is no device that measures signs \emph{without making such arbitrary conventions}.  These gauge theories ensure that we don't arbitrarily assign absolute signs to charges.  The other `charges' in the standard model and string theory are more exotic but the general theme remains the same.

There is, of course, some sense to the idea that it is convenient to make arbitrary conventions.  When measuring distances while designing a house it is convenient to be able to measure them accurately with respect to a certain reference point.  (Clearly, however, it is not convenient to always measure distances from the Eiffel tower.)  So we come again to the question of whether we should simply use arbitrary conventions \emph{because} they are convenient?  The simple fact is that we \emph{should} make arbitrary conventions for convenience but only when we \emph{do not care how such `positions' or `signs' come about}.  Arbitrary conventions do not have any direct explanatory power at all, they are tools which we use to explain \emph{other} things.  The kinematics of space allow us to explain how to build a house but do not help us explain how the particular space in which the house is built came about.  Also, we are free to make arbitrary conventions while discussing buildings, but we are not so free when discussing the whole universe as we have no reference points outside the universe (we must define distances with respect to things inside, so we may as well, all things being equal, define them relationally).

Clearly we do not claim here that all these theories are \emph{uniquely} relational.  We could interpret anything using arbitrary conventions when it is useful to do so.  We simply mean that in \emph{designing} theories which have explanatory power we should not introduce arbitrary conventions and that it is, in fact, pragmatic and convenient not to introduce such things.  We are far more likely to find results that aren't factitious by ensuring we don't use factitious assumptions.

To explain what we mean when we say that arbitrary conventions do not have any direct explanatory power at all it is perhaps easier to turn to yet another of the great modern relational theories; namely Darwinian evolution (which is clearly a \emph{physical} theory even if also a biological one).  We can easily discuss the `kinematical' distinctions between different species alive today.  Species can be said to be `alike' if they share common features.  Of course the notion of `species' is a rather arbitrary convention, as is the notion of `common features' but, nonetheless, many of us desire an explanation for why those species that are alike are, in fact, alike.  Clearly we do not have a `speciesometer' which reads off which species an organism belongs to without making arbitrary conventions---species, like signs and positions, are themselves convenient conventions!  Hence it is not the conventions that matter so much as the relationships between them.  Thus we require a dynamical explanation for how the particular relationships between the present set of species came to be.  This is what evolution gives us; thus, we argue, evolution is a relational theory.  It is a general feature of all relational theories that they explain what were previously considered `kinematical' or absolute structures by dynamical means.  In GR it is the causal structure that is considered dynamical (whereas in Newtonian physics it was considered a kinematical idea).  The present particular causal structure of the universe is something that is dynamically contingent upon the distribution of matter and energy contained within it and the history of the universe.  Similarly, the taxonomic relationships between species, previously considered `kinematically', are now explained dynamically.  Such relationships are contingent upon the particular history of the \emph{process} of life on Earth. (There is a similar explanatory power in gauge theories related to what is called `spontaneous symmetry breaking' but we won't go into that here).  Furthermore, such relationships \emph{ought} to be explained in such a manner---we explicitly have no other way to explain them.  We cannot explain a `thing' without explaining how that `thing' came to be.

So, this is all well and good, but such an account of modern physical theory is surely uncontroversial?  Why do we feel the need to write about this?  Note, however, there is one of the great modern theories which is emphatically \emph{not} designed in a relational manner; namely quantum theory.  We want to bring quantum theory into the fold (for other attempts see \cite{Marlow06c,Rovel96,SR06}).

Please excuse the further ambulation; we shall discuss quantum theory shortly but we must first tackle quantum theory's classical analogue: probability theory.  Here we have a situation which is clearly analogous to the theory of evolution; we have a set of convenient conventions (probabilities) which need explanation.  In order to follow a relational approach the things that matter are clearly the relationships between probabilities and not the probabilities themselves.  To borrow Mana's phrase \cite{Mana04}, probabilities are not ``experimentally observed regularities''; just as positions, signs and species are not.  Rather they are `theoretically \emph{assigned} regularities'.  So, if we are to also bring probability theory into the fold then we must explicitly discuss it in relational terms.  The only approach we know of that vaguely does this, at least in part, is Cox's approach to probability theory \cite{CoxBOOK,JaynesBOOK}.  Here the relationships which matter are the \emph{functional} relationships between probabilities.  In evolutionary terms we have a certain amount of freedom in how we assign species to organisms;  similarly, in probability theory, we have a certain amount of freedom in assigning probabilities to propositions (because they are arbitrary conventions)---there is no \emph{a priori} reason we should use real numbers, nor, if we do use the reals, to use the range $[0,1]$ in particular.  We must, however, ensure that the relevant relationships between probabilities are maintained regardless of whether one uses a particular convention and this is what Cox's axioms of probability theory ensure.  But what are these axioms?

Let's discuss a set of propositions $\{\alpha, \beta, \gamma,...\}$ which can be related by using the standard `and', `or' and `not' operations, denoted $\cap$, $\cup$ and $\sim$ respectively.  Such operations behave as one expects na\"{\i}vely---think of Venn diagrams or Boolean logic.  Cox's two axioms are as follows:

\begin{equation}
p(\alpha \cap \beta \vert I) := F[p(\alpha \vert \beta I), p(\beta \vert I)]
\label{COX1}
\end{equation}

\noindent and

\begin{equation}
p(\sim \alpha \vert I) := G[p(\alpha \vert I)],
\label{COX2}
\end{equation}

\noindent where $F$ and $G$ are two arbitrary functions that are sufficiently well-behaved for our purposes \cite{JaynesBOOKb9} and $I$ is the notation we use to emphasise that probabilities are contingent upon prior information.  We do not need an $\cup$-axiom because the operations $\cap$, $\cup$ and $\sim$ are interdependent.  So we argue that it is rational to presume that the probability that we assign to $a$ must, in the least, be functionally related to the probability we assign to $\sim a$---this is (\ref{COX2}), one of the minimal desiderata of rationality that Cox used.  We call it minimal because all known probability theories obey it, and we can, it seems, all agree that any probability theory ought to obey it. If a probability theory were not to obey it then $p(a \vert I)$ would be unrelated to $p(\sim a \vert I)$ and such $p$'s would not be useful as any notion of `preference' or `plausibility' at all.  The other axiom (\ref{COX1}) is a little more complicated, but it certainly seems natural that in order to assign a probability to the proposition $a \cap b$ we should not need any more information than the probability that we assign to `$a$ given $b$ is the case' and the probability we assign to $b$ being the case.  Equally, by swapping $a$'s for $b$'s and $b$'s for $a$'s we have an analogous conclusion.  All known probability theories satisfy these axioms and they are reasonable (we make very minimal assumptions about how some probabilities are related to others).  Of course, there is nothing absolute about rationality and some reasoner of the future might be able to argue reasonably that we should use other axioms or introduce further ones.  The important point is not that we regard these axioms as inherently `rational' (as we don't) but rather that we have no clear and unambiguous argument for using axioms that go beyond these (such as demanding \emph{a priori} that probabilities are real numbers).  We don't have a `probability-computer' to which we can feed-in any proposition (and that proposition alone) such that it calculates the `absolute' real probability unique to that proposition.  Rather probabilities are defined relationally over the whole algebra of propositions using Cox's axioms.  If you accept the argument for defining positions relationally, an analogous argument follows through for probabilities.

We say that $F$ and $G$ are `arbitrary' in the sense that we can find no compelling reason why we ought to use a particular $F$ and $G$.  This idiosyncratic `relational' way for putting it is, of course, our own---as is any error made in doing such a thing---but we think that this explains Cox's approach quite well.  Cox explicitly also assumed that probabilities are real numbers which is an assumption above and beyond (\ref{COX1}) and (\ref{COX2})---we call such an assumption Cox's zeroth axiom.  Implicit as part of this zeroth axiom is the minimal `monoticity' assumption:

\begin{equation}
\mbox{if } a \leq b \mbox{ then, presumably, } p(a \vert I) \leq p(b \vert I),
\end{equation}

\noindent where $\leq$ is, in the least, a partial order in both the context of the Boolean algebra of propositions and the probability space.  For example, in a Boolean logic represented by a Venn diagram `$a \leq b$' is represented by set inclusion. (To avoid confusion one might prefer to use different symbols for the partial order on the proposition space and the partial order on the probability space but we will soon run out of convenient symbols.  The meaning of the symbol $\leq$ should be evident by context.)  Standard probability theory obey's all these axioms, but it does not satisfy these axioms uniquely.  And this is exactly the point; the above axioms seem reasonable for \emph{any} probability theory.  If we were to introduce further axioms that we are not rationally compelled to use (in order to arbitrarily ensure that probabilities were real numbers for example) then we would be disobeying the PSR.

So we can categorise the essential relationships between probabilities without making particular arbitrary conventions (\emph{i.e.} we don't choose a particular $F$ or $G$ and nor do we choose to use real numbers \emph{per se}).  This is analogous to framing general relativity in a `co-ordinate free' or `background independent' way (for an accessible introduction to background independence see \cite{Smolin05}).  All relational theories can be framed in such a `background independent' manner and there is a certain beauty in doing so since we explicitly remove the arbitrary conventions from the very formalism we use---this increases our chance of speaking sensibly since we have removed all factitious conventions from the very mathematical language we use to speak about the theory\footnote{\emph{Cf.} Wittgenstein's famous proposition 7: ``What we cannot speak about we must pass over in silence'' \cite{WittgenBOOK}.}. Thus we do not need to worry about any odd factitious relationships that arise \emph{through} the use of odd factitious conventions (\emph{cf.} \cite{Marlow06c,SR06,Marlow06d,JaynesBOOKsub}).

Let us reiterate that we are not claiming that these relational theories are \emph{absolutely} relational in their outlook---there is no such thing as absolute rationality.  For example, in GR the topology remains kinematically fixed and is kept as part of the `background' of the theory \cite{Smolin05}.  Similarly, in our probability analysis one might ask the question ``why assume these relationships are \emph{functional} relationships?''.  Perhaps there are some hidden assumptions in the use of functions over more abstract and general notions (perhaps involving category theory).  This is where the competing form of pragmatism kicks in.  The theory would be completely unwieldy to handle if we tried to go too far too quickly.  Nonetheless, the fixed background topology is beginning to be seriously questioned in quantum gravity theories, and one might later question the use of functions \emph{per se} in this probability analysis.  Strictly speaking, we find partially relational theories and then try to make them \emph{more} relational.  Nor do we claim that standard probability theory (which uses further dubious axioms to ensure that probabilities are real numbers) isn't useful, it clearly is; nonetheless, there might be cases where using standard probability theory causes factitious inferences which result from the use of factitious axioms (axioms which don't obey the PSR).

So, we can frame probability theory in a manner that arguably obeys Leibniz's rational desiderata such that we remove arbitrary conventions from the very formalism we use.  Can we extend this relational approach to probability theory to give a \emph{dynamical} explanation of the catalogue of relationships between probabilities?  How does this catalogue of relationships between probabilities evolve?  In GR we can show that the causal relationships between spacetime points are dynamically contingent relationships.  Similarly for the taxonomic relationships we find in biology.  What is the analogous dynamical explanation we need for probability theory?

The key is in our gnomic use of the `context' $I$ in our notation.  Clearly we can use other propositions as such `contexts'---we can ask: ``What is the probability of proposition $\alpha$ being true \emph{given} that $\beta$ is true?'' and we denote such contingent probabilities as $p(\alpha \vert \beta I)$.

Cox's $\cap$-axiom (\ref{COX1}), as Cox proved \cite{CoxBOOK}, ensures that Bayes' rule is satisfied:

\begin{equation}
p(\alpha \vert \beta I) = \frac{p(\beta \vert \alpha I) p(\alpha \vert I)}{p(\beta \vert I)}.
\end{equation}

\noindent  Hence we can `update' probability assignments.  Probabilities are contingent on the particular propositions that happen to be known to be true.  But are they \emph{dynamically} contingent?  Of course they are as long as one assumes that propositions become apt \emph{dynamically}.  Hence the functional relationships between probabilities are not fixed, absolute or `kinematical' relationships but are dynamically contingent upon which propositions are known to be true (and other prior information).  We presume that there is a dynamical process by which some propositions become apt and we argue that it would be irrational to think otherwise.  In standard probability theory we usually presume that there \emph{is} such a dynamical process and that it is described by classical physics.  A proposition like ``the rocket lands near me on the Earth's surface when the time on my watch is between $t_1$ and $t_2$'' is a proposition that, if it becomes true at all, will become true dynamically.  Hence in classical probability theory the functional relationships between probabilities will be dynamically contingent relationships by Bayes' rule.

We keep the $I$ in the notation because we don't know how to interpret a probability that \emph{isn't} contingent as then they would be `kinematical' or absolute notions---something given by a fictitious `probability-computer'.

So, finally, we come to quantum theory.  Quantum theory is tautologically a generalised form of probability theory so perhaps this rationalist account will help us.

In probability theory the natural algebra of propositions is Boolean.  We peg probabilities to these propositions in a manner which maintains certain functional relationships between probabilities.  In quantum theory we have a slightly different set of propositions that behave in a slightly different way.  Also, in quantum theory we don't, strictly speaking, peg probabilities to these propositions.  Rather we peg something else which we shall, naturally, call a `peg'.

So in quantum theory we have a natural set of propositions which we again call $\alpha, \beta, \gamma$, and so forth, since they have the same meaning as they do normally.  They are simply statements like ``it is true that the value of quantity $A$ is $a$''.  However, these propositions behave in an odd way---they aren't related by the normal Boolean operations $\cap$, $\cup$ and $\sim$.  Rather there are novel, but still \emph{roughly} analogous, operations $\wedge$, $\vee$ and $\neg$ (and a notion of partial order that we call again $\leq$).  Thus, for example, the `and' operation $\wedge$ allows us to make propositions like ``it is true that the value of quantity $A$ is $a$ `and' it is true that the value of quantity $B$ is $b$''.  These operations aren't the same as the `normal' Boolean ones, but they are sufficiently similar for us to make na\"{\i}ve progress. If we were to peg truth values then we would assign either $0$ or $1$ to each of these propositions in a way which satisfies certain consistency conditions or, if we were to peg `probabilities' to these propositions, we would assign real numbers in the interval $[0,1]$ to all these propositions in a way that ensures the relationships between `probabilities' are maintained.  It turns out we cannot consistently do the former\footnote{This is, for those in the know, the Bell-Kochen-Specker theorem, and is related to the problem of using Boolean logic to assign Boolean truth values to a quantum one,  but it won't concern us here.} but we can do the latter using what is called Gleason's theorem \cite{Gleason57}.  These `probabilities' don't behave like normal probabilities in probability theory---this causes a lot of confusion and it is not clear whether this confusion arises simply because we tend to think of the probabilities we assign to Boolean propositions as the same kinds of arbitrary conventions we assign to quantum propositions.  Hence we would prefer to use a different name.  For the time being we resort to calling these conventions `pegs' as this seems to emphasise what we are doing---we peg `pegs' to propositions\footnote{Similarly, we `posit' positions in Newtonian physics.}---and we don't have to change the notation because `peg' begins with a `p'.  Standard probabilities are thus a type of peg which is relevant for Boolean propositions.  In order to investigate whether some \emph{other} type of peg\footnote{Emphatically not probabilities \emph{per se}.} is relevant for quantum theory we approach the problem from a relational perspective.

We can, for example, na\"{\i}vely use Cox's axioms as we did above:

\begin{equation}
p(\alpha \wedge \beta \vert I) := F'[p(\alpha \vert \beta I), p(\beta \vert I)]
\label{PEG1}
\end{equation}

\noindent and

\begin{equation}
p(\neg \alpha \vert I) := G'[p(\alpha \vert I)]
\label{PEG2}
\end{equation}

\noindent where $F'$ and $G'$ are again arbitrary functions that are sufficiently well-behaved for our purposes.  Similarly we can also assume a minimal version of Cox's zeroth axiom such that

\begin{equation}
\mbox{if } \alpha \leq \beta \mbox{ then, presumably, } p(\alpha \vert I) \leq p(\beta \vert I).
\label{PEG0}
\end{equation}

We have called conventions which satisfy these axioms `pegs' rather than `probabilities' simply to emphasise that they are things that we append to propositions and to ensure we don't confuse them with `probabilities' \emph{per se} which are pegs pegged to a Boolean logic and are often additionally constrained to be real numbers in the interval $[0,1]$.  Thus we need not be so confused when pegs behave differently when pegged to propositions which, in turn, obey a different logic.  Now, in standard probability theory we were happy to presume that there is a dynamical reason why some propositions become apt.  Thus we could explain the relationships between probabilities by showing that they are dynamically contingent upon a given realistic kind of evolution. Before we had classical physics to hand in order to explain the dynamical contingency of relationships between probabilities.  In quantum theory we don't yet understand the underlying dynamical physics---this is perhaps what people mean when they say that ``no-one understands quantum theory''.

Some people have attempted to make predictions about the underlying physics---that it is somehow `nonlocal' or `holistic' and so forth, but such ideas are, at their worst, waxed philosophy or, at their best, not wholly unconvincing \cite{EPR,BellBOOK}.  For example, `holism' is simply an uninteresting tautology---the whole is more than the some of its parts simply because the parts \emph{interact} (or because `parts' are \emph{defined} by the unique catalogue of their relationships with all other `parts').  Of course, if we assume that quanta buzz around as if they behaved classically then we get issues with `nonlocal causality' but there is nothing to stop quantum theory being causally local when treated on its own terms (so that we could even explain it to an alien).  For example, Bell's \cite{BellBOOK} definition of locality relies on the mere assumption that the peg assigned to the proposition for an event to be the case should remain unchanged when we update it, using Bayes' rule, to be contingent upon the outcome of an experiment at spacelike separation.  This assumption is disobeyed by quantum theory in the orthodox interpretation and many people then call quantum theory `nonlocal' because of this fact.  In GR, the very definition of what we `mean' by the position of a particle on Earth relies on its relationships with the background stars regardless of the fact they are far away.  We don't call this property `nonlocality' in GR (for good reason, it is a causally local theory), and nor should we call the analogous property in quantum theory or probability theory `nonlocality' either.  In a relational approach, part of the \emph{very definition} of what we mean by a `probability' or a `peg' relies on functional relationships between propositions (regardless of whether the events to which those propositions refer are causally separate). Similarly, in evolutionary terms, geologically isolated species might share a common phenotype which cannot be explained by their common ancestry but, nonetheless, these seemingly `nonlocal' relationships can be explained to be factitious relationships through evolutionary theory---they might be `convergent' species (\emph{i.e.} we need not postulate that one of these isolated species evolved from the other). We are not rationally compelled to assume that there are absolute species, positions or probabilities.  If we simply postulate that seemingly `nonlocal' relationships are \emph{factitious} then we are freed to search for a rational account of them---one in which we will only accept causal nonlocality if we find some physical entity (or perhaps information) which travels at faster than the speed of light (by the PSR).

Note that it might be that we are making a category error by maintaining these \emph{particular} peg relationships in quantum theory---a less archaic peg taxonomy might be more apt.  There might be other criteria of rationality that we `ought' to use; this would involve either changing the peg axioms---(\ref{PEG1}), (\ref{PEG2}) and (\ref{PEG0})---or by using a slightly different quantum propositional algebra;  Isham and Butterfield have investigated an approach of the former variety in \cite{IB98} and we have investigated a theory of the latter in \cite{Marlow06c,Marlow06b}.

In conclusion, relationalism provides a way to be parsimonious about the foundations of physical theory.  Such relationalism has yet to be applied fully to quantum theory, and might yet provide significant physical insight.

\subsection*{Acknowledgements}

Thanks are due to EPSRC for funding this work.


\begin{thebibliography}{99}
\bibitem{unknown}  This pedagogical explanation using `positometers' is not new.  We have read it somewhere, though unfortunately we cannot remember where.
\end{thebibliography}
\end{document}